\documentclass[12pt]{iopart}
\usepackage{epsfig}
\begin{document}

\title{Field emission theory beyond WKB - the full image problem}
\author{ T C Choy, A H Harker and A M Stoneham}
\address{Department of Physics and Astronomy, University College London,
Gower Street, London WC1E 6BT}

\ead{a.harker@ucl.ac.uk}

\begin{abstract}
The classic theory of electron field emission from a cold metal
surface due to Fowler and Nordheim (FN) is re-examined and found
to violate the validity criteria for the WKB approximation, for
electric fields greater than about $1~{\rm V}/\mu{\rm m}$. In this
study we shall examine the complete solution without invoking the
WKB approximation in order to assess the reliability of the FN
theory as widely used for the interpretation of experimental data.
Particular problems occur when the barrier height (and therefore
indirectly also the width) is significantly reduced by the image
or some external effects. Further refinement of the theory will be
discussed by considering the effects of screening, which can be
one mechanism for the barrier height reduction, in addition to the
widely known negative affinity of diamond like carbon systems. A
comparison with experimental data from carbon field emitters shows
that the enhanced current found in this paper may provide an
explanation for strong field emission observed recently in carbon
based samples.
\end{abstract}
\submitto{JPCM} 
\maketitle
\maketitle
\section*{Introduction}
\label{Introduction}
 The classic theory of electron field emission
from a cold metal surface due to Fowler and Nordheim (FN)
\cite{FowlerN,Nordheim,GoodMuller} was one of the earliest
applications of wave mechanics since its foundation in the early
1920's. With increased interest in field emission processes,
especially for use in display screens \cite{PFE}, it is worth
examining closely the foundations of this theory. As well as
metal-vacuum interfaces, proposed applications of the theory
includes graphite-vacuum, graphite-insulator-vacuum,
carbon-nanotube-vacuum and Schottky junctions \cite{Seitz}. Since
our approach can be generalized to other cases of interfaces
between dissimilar media, it may have application to the systems
discussed by Mott and Gurney \cite{MottGurney} and Herring
\cite{Herring}.  These other cases might include vacuum breakdown
and electrical forming processes \cite{Dearnaley}.  In section
\ref{Experiments} we show that our new analysis removes
discrepancies of many orders of magnitude between theory and
experiment.

 In the FN theory, the role of the Fermi surface,
Fermi-Dirac statistics and in particular the unique property  of
quantum mechanical tunnelling were key concepts. The essentially
equilibrium distribution of electrons is modeled by a one
dimensional ($x$ dependent only) Hamiltonian:
\begin{equation}
H_0 = \frac{p_x^2}{2m} + V(x), \label{Hamiltonian}
\end{equation}
where all energies are measured with the same reference as the
effective potential $V(x)$. Then the current density $J$ of the
emitted electrons is given by an integral over the energy $W$:
\begin{equation}
J = e \int_{-W_a}^{\infty} P(W) dW ,\label{Current}
\end{equation}
where $P(W) dW$ gives the number of electrons within $dW$ that
emerge form the metal per second per unit area, with $W_a$\ an
appropriate lower energy reference. The quantity $P(W)$  is easily
expressed as the product of a quantum mechanical tunnelling
transmission coefficient $T(W)$ and a Fermi type supply function
$N(W)$:
\begin{equation}
N(W) = \frac{4\pi mkT}{h^3} \ln \left \lbrace {1+ \exp\left [{
-(\frac{W-\epsilon_F}{kT})}\right ]}\right \rbrace ,
\label{SupplyFunction}
\end{equation}
in which $\epsilon_F$ is the Fermi energy. In view of the
complexity of the Schrodinger equation, the transmission problem
was originally treated within a WKB approximation or variants of
it \cite{GoodMiller}:
\begin{equation}
T(W) = \exp \left [{ - \int_{x_1}^{x_2} \left ( {
\frac{8m}{\hbar^2}(V(x)-W)} \right )^{1/2}dx}\right ]   ,
\label{WKBTransmission}
\end{equation}
in which $x_1$\ and $x_2$\ are the classical turning points of the
potential $V(x)$. It is a well known result however that the
validity of the WKB approximation involves a criterion, see
section \ref{WKBValidity}, that can be derived from the
requirement that the variation of the deBroglie wavelength of the
electron must be less than the dimension of the region of
tunnelling \cite{LandauLifshitz}.  This restriction on the
validity of the FN theory was unfortunately not discussed in many
of the earlier papers, including the classic paper of
\cite{GoodMuller}. The limitations of the WKB approach are well
known, however, in the treatment of thermionic emission; a common
technique for improving the calculation\cite{GuthMullin} is to
make a parabolic approximation to the top of of the barrier.
Nevertheless, FN theory has been applied in regimes where it is
not valid. In section \ref{ImageFreeCase} we shall examine the
image-free problem exactly \footnote{Since image effects are less
important for silicon based  Schottky junctions due to the large
dielectric constant $\epsilon_s \approx 11.9$, these results are
actually very relevant to the latter, which however we shall not
take up here in this paper.}, whose solution in one dimension is
amenable to analytical treatment in terms of Airy type functions.
Some preliminary assessment of the FN-WKB can already be made for
this case, which is a preamble to the full image problem which we
shall treat in section \ref{WithImage} by various analytical
methods. Interestingly the tunnelling problem close to the barrier
top appears to be amenable to analytical treatment as well and has
not been published to the best of our knowledge. We shall note
that the failure of the WKB criteria coincides quite closely to
this regime, hence the availability of analytical results are
extremely welcomed. Next we shall note that the role of screening
has also been omitted in the original FN theory. We shall
introduce the additional screening potentials in section
\ref{Screening} and thereby propose that this is one mechanism for
a significant barrier reduction, in addition to negative electron
affinity attributed to diamond like carbon systems. In section
\ref{Experiments} we shall apply our theory to the analysis of
experimental data for carbon field emitters\cite{TuckLatham}.
Regardless of reasonable assumptions of point tip field
enhancement effects, we shall see that the FN theory does not even
get close to the order of magnitude agreement with the data by
comparison with ours.  We shall conclude in section
\ref{conclusion} with a summary of our results, postponing further
solutions of the problem involving screening to a future paper.
\section{WKB Validity criteria} \label{WKBValidity} Following
\cite{LandauLifshitz} it is straightforward to derive the
criterion for the validity of WKB:
\begin{equation}
p^3 >> m \hbar eF, \label{WKBCriteria1}
\end{equation}
where $F$ is the external field and we shall ignore the electron
image for the moment. Here $p$ is the electron momentum so that:
\begin{equation}
(2m(\phi-eFx))^{3/2} >> m \hbar eF, \label{WKBCriteria2}
\end{equation}
where $\phi$ is the work function. Assuming a carrier effective
mass equal to the free electron mass and inserting numerical
values appropriate to a field in V/$\mu$m and a work function in
eV, (which shall be our units throughout this paper) gives
\begin{equation}
 \zeta (F,\phi) = 0.046 {F^{1/3}\over (\phi)^{1/2}}<<1, \label{WKBCriteria3}
\end{equation}
in which the tunnelling length $x$ has been set to zero,
representing the best case for WKB, the very top of the barrier
($x$ is typically of the order of nm). Typical numerical values
for $\zeta$ are given table~\ref{tab_1}, where we see that the WKB
approximation begins to be questionable for fields above about $10
{\rm V}/\mu{\rm m}$ for typical work function values and is
certainly seriously questionable in the range close to dielectric
breakdown fields which are typically $10^3~{\rm V}/{\rm\mu m}$ . A
smaller effective mass such as for p-Si where ($m^*=0.6$ for
holes) would increase $\zeta$ further. Later on in the paper we
shall show further anomalies with the FN theory when we are close
to this limit of validity.  In order to proceed and be able to
assess the approximation further we shall have to resort to a full
solution of the Schrodinger equation which we shall begin in the
next section by first ignoring the image potential.
\begin{table}
\caption{\label{tab_1} Values of the function $\zeta(F,\phi)$. For
WKB theory to be valid, $\zeta(F,\phi)<<1$.}
\begin{indented}
\item[]\begin{tabular}{@{}llll} \br $F({\rm V}/{\rm\mu
m})$&$\phi = 1
{~\rm e\rm V}$&$\phi = 5 {~\rm e\rm V}$&$\phi = 10 {~\rm e\rm V}$\\
\mr
1    & 0.046  & 0.021 & 0.015 \\
10   & 0.099  & 0.044 & 0.031 \\
100  & 0.213  & 0.096 & 0.068 \\
1000 & 0.460  & 0.206 & 0.146 \\
\br
\end{tabular}
\end{indented}
\end{table}
\section{Exact solution without electron images}
\label{ImageFreeCase} We shall start with the image free
tunnelling problem given by the Schrodinger equation:
\begin{equation}
\psi^{\prime\prime}(x) + \frac{2m}{\hbar^2}[W-V(x)]\psi(x)=0,
\label{Schrodinger1}
\end{equation}
where $W$ is the energy eigenvalue of the Hamiltonian $H_0$ with
the potential given by \cite{GoodMuller}:
\begin{equation}
V(x) =\cases{ -W_a\quad  &if $x<0$ -- region 1; \cr
              V_0 - e F x &if $x>0$  -- region 2.\cr
              }
\label{ImageFreePotential}
\end{equation}
Here $V_0$ under normal circumstances is zero which yields the
work function $\phi$ as the barrier height and $F$ is as before
the applied external field. For later convenience we shall recast
Eq(\ref{Schrodinger1}) into the more compact form:
\begin{equation}
\psi^{\prime\prime}(x) + (\epsilon+\alpha x)\psi(x)=0,
\label{Schrodinger2}
\end{equation}
with
\begin{eqnarray}
\alpha & = &\frac{2meF}{\hbar^2} \approx  F \times 2.55989 \times
10^7 ({\rm\mu  m})^{-3},
\nonumber \\
\epsilon & = & \frac{2m(W-V_0)}{\hbar^2}  \approx (W-V_0) \times
2.55989 \times 10^7 ({\rm\mu  m})^{-2}, \label{constants}
\end{eqnarray}
in which $F$ is measured in ${\rm V}/{\rm\mu m}$.  To the best of
the authors' knowledge this elementary quantum exercise does not
seem to have been fully documented in the literature, although
variants of it have been scattered around \cite{Roy, Delbourgo,
Duke, Moll}. Following Landau and Lifshitz \cite{LandauLifshitz},
we shall transform to the new variable $\xi =
(x+\epsilon/\alpha)\alpha^{1/3}$. Then equation
(\ref{Schrodinger2}) in region 2 becomes the typical Airy's
equation \footnote{The advantage of using the Airy function
formulation is that it naturally takes care of the long range
boundary condition posed by the electric field.}:
\begin{equation}
\psi^{\prime\prime}(\xi) + \xi \psi(\xi)=0. \label{Schrodinger3}
\end{equation}
However some care needs to be exercised in selecting the
appropriate type of Airy functions, since the standard Ai function
has asymptotic form \cite{LandauLifshitz, Watson}:
\begin{equation}
\psi(\xi)={\rm{Ai}}(-\xi)  \approx \xi^{-1/4} \sin(\frac{2}{3}
\xi^{3/2}+ \pi/4). \label{AsymptoticAi}
\end{equation}
As this also contains a reflected wave, it is not appropriate to
our boundary conditions. The correct approach is to revert to the
original 1/3 fractional Bessel functions definition for the Airy
functions and then construct appropriate linear combinations to
yield the forward travelling wave solution only. This approach of
course leads to a modified appropriate Hankel function which as we
shall see has the correct asymptotic form required. It is this
construction that seems to be rather rarely seen in the literature
except perhaps in the WKB derivation of Moll \cite{Moll} . With
this proviso, we can write down the solution for the wave
functions of equation (\ref{Schrodinger3}) as:
\begin{equation}
\psi(x) = \cases{e^{ikx}+ r e^{-ikx} &in region 1; \cr
                  t[u_1(\xi)-e^{-i\pi/3}u_2(\xi)] &in region
                  2,\cr}
\label{wavefunctions1}
\end{equation}
where $k=(2m(W+W_a)/\hbar^2)^{1/2}$ and the $u_1$ and $u_2$ are
given by Bessel functions of order $1/3$, here defined as:
\begin{equation}
u_1(\xi)  =\frac{1}{3} (\pi\xi)^{1/2}
J_{-1/3}(\frac{2}{3}\xi^{3/2}),
\label{Besselfunction1}
\end{equation}
and
\begin{equation}
u_2(\xi)  =\frac{1}{3} (\pi\xi)^{1/2}
J_{1/3}(\frac{2}{3}\xi^{3/2}), \label{Besselfunction2}
\end{equation}
respectively. Note however that analytical continuation to
negative $\xi$ requires some care: see Appendix A. It is then
straightforward to obtain the reflection and transmission
amplitudes by matching the wave functions and their derivatives at
the interface $x=0$. We provide the results here in terms of
$H_{1/3}=u_1-e^{-i\pi/3}u_2$ which is a slightly modified Hankel
function
\begin{eqnarray}
t &=&  \frac{2}{H_{1/3}(\xi_0)-ik_0^{1/3}H^\prime_{1/3}(\xi_0)} \nonumber \\
r &=&
\frac{H_{1/3}(\xi_0)+ik_0^{1/3}H^\prime_{1/3}(\xi_0)}{H_{1/3}
(\xi_0)-ik_0^{1/3}H^\prime_{1/3}(\xi_0)}
, \label{Amplitudes}
\end{eqnarray}
where $\xi_0=\xi(0)=\bar\epsilon=\epsilon/\alpha^{2/3}$ while
$k_0^{1/3}=\alpha^{1/3}/k$.  An important check must be made to
these formulas which requires that the transmission coefficient:
\begin{equation}
T=\frac{k_0^{1/3}}{4}|t|^2, \label{TransmissionCefficient}
\end{equation} and reflection coefficient: $R=|r|^2$ satisfy the
unitarity property $T+R=1$. This is easily checked to be the case
when appropriate use is made of the Wronskian identity:
\begin{equation}
u_1u_2^{\prime}-u_2u_1^{\prime}=\frac{1}{2\sqrt{3}}.
\label{Wronskian}
\end{equation}
For sufficiently small fields, we can obtain an asymptotic
expansion (where $\xi_0 \rightarrow \infty$ and $k_0 \rightarrow
0$) that yields:
\begin{equation}
T_{asymp} \approx
\frac{4k_0^{1/3}|\xi_0|^{1/2}}{(1+k_0^{2/3}|\xi_0|)}\
e^{-\frac{4}{3}\xi_0^{3/2}}, \label{AsympLimit1}
\end{equation}
the prefactor being a new result not obtainable within WKB. For
convenience we set $W_a=2\phi$ which for $\phi$ of order $1{~\rm
e\rm V}$ is adequate as a lower bound (at room temperatures), the
exception being very narrow band semiconductors which we are not
concerned with here. We shall at the moment ignore any possible
barrier height lowering (i.e. image free case) so that $V_0=0$ is
the vacuum level and consider tunnelling near the Fermi level, so
that $W=-\phi$ and thus $k_0^{2/3}|\xi_0|=1$ . Then we see that
the above prefactor differs from the WKB result by a mere factor
of two i.e.:
\begin{equation}
T_{asymp} \approx 2\ e^{-\frac{4}{3}\xi_0^{3/2}}.
\label{AsympLimit2}
\end{equation}
That this asymptotic result is valid for only small fields is
clearly seen now for otherwise if we naively allow the field to be
large, then we could arrive at an unphysical result in which $T>1$
which is absurd. In fact this occurs for $\phi=1{~\rm e\rm V}$
when $F$ reaches the order of $9.9~{\rm kV}/{\rm\mu m}$. Hence the
use of the WKB formula for large transmissions without knowledge
of the prefactor in the transmission can be dangerous.
Unfortunately this caution does not seem to have been exercised by
Good and M\"uller \cite{GoodMuller}, nor by Murphy and Good
\cite{MurphyGood}. The latter have derived widely used criteria
\cite{Stratton} for the validity of FN theory on the basis of the
WKB integrand properties equation (\ref{WKBTransmission}) alone,
which is thus a logically inadequate procedure. If we examine the
classic FN formula for the emission current:
\begin{equation}
\fl J=\frac{1.54\times 10^{2} F^2}{\phi\ t^2(0.0379\times
F^{1/2}/\phi)}\ \exp-(\frac{6830\times
\phi^{3/2}}{F}\upsilon(0.0379\times F^{1/2}/\phi)) ~{\rm A~\rm
{cm}^{-2}} \label{FNcurrent}
\end{equation}
in which $t(x)$ and $\upsilon(x)$ are given in terms of elliptic
integrals, for $\phi=2{~\rm e\rm V}$ say, the exponent practically
vanishes at $F\approx 3000{\rm V}/{\rm\mu m}$. Hence the entire
emission current comes from the supply function which is the $F^2$
factor and, as discussed above, this can be untenable. In fact it
is well known from elementary quantum mechanics that the
transmission coefficient is in general not unity at the top of the
barrier \cite{Schiff}.

\begin{figure}
\begin{center}
\includegraphics[scale=1.0]{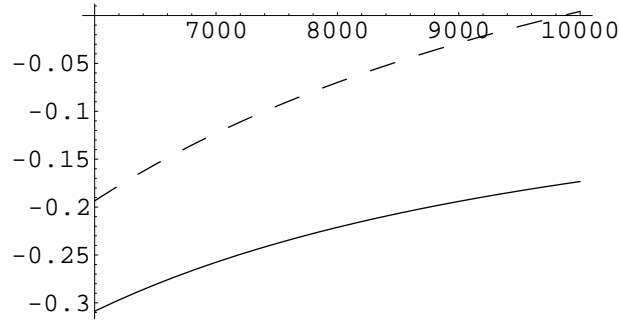} \end{center} \caption{\label{fig 1}Log-linear
plot of the exact transmission coefficient (solid line) equation
(\ref{TransmissionCefficient}) vs the asymptotic formula equation
(\ref{AsympLimit2}) (dashed line) close to the Fermi-level in the
image-free case for large fields.  $F$ is in units of $\rm V/\mu
m$.}
\end{figure}

\begin{figure}
\begin{center}
\includegraphics[scale=1.0]{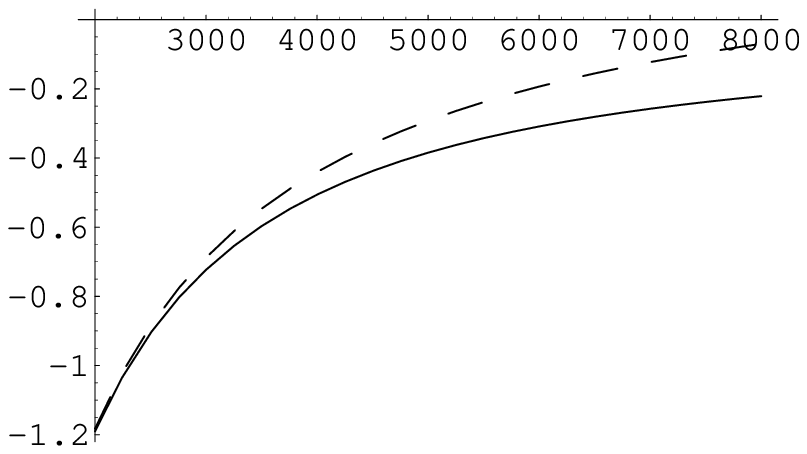} \end{center} \caption{\label{fig 2}Log-linear
plot of the exact transmission coefficient (solid line) equation
(\ref{TransmissionCefficient}) vs the asymptotic formula (dashed
line) equation (\ref{AsympLimit2}) close to the Fermi-level in the
image-free case for smaller fields. $F$ is in units of $\rm V/\mu
m$.}
\end{figure}

\begin{figure}
\begin{center}
\includegraphics[scale=1.0]{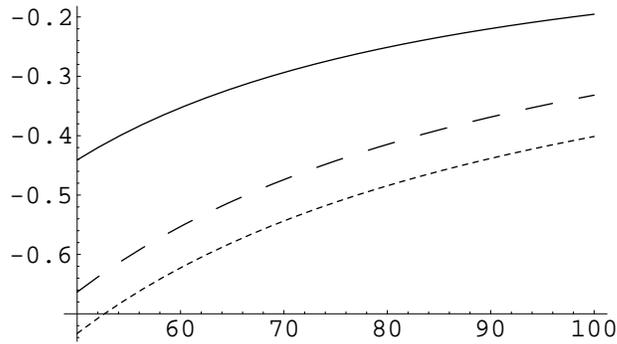} \end{center} \caption{\label{fig 3}Log-linear
plot of the exact transmission coefficient (solid line) equation
(\ref{TransmissionCefficient}) vs the asymptotic formula (dashed
line) equation (\ref{AsympLimit1}) and vs the standard WKB formula
(dotted line) close to the top of the barrier ($V_0=-0.95~\phi$)
in the image-free case. $F$ is in units of $\rm V/\mu m$.}
\end{figure}

Later on, in section \ref{WithImage}, we shall see that an exact
solution is obtainable for the full image theory at the top of the
barrier, in which the transmission coefficient has a finite value
less than unity, but nevertheless with significant tunnelling
current contributions. As far as we are aware, this solution has
not been noted in the field emission literature for the last 40
years. In figure \ref{fig 1} we provide two curves that compare
the exact transmission coefficient equation
(\ref{TransmissionCefficient}) for the image free case against the
asymptotic formula equation (\ref{AsympLimit2}) close to the Fermi
level for very large fields where the latter becomes problematic.
Fortunately these fields are very large and being much greater
than typical breakdown fields this failure is unimportant in
practice. We note that for smaller fields figure \ref{fig 2} the
WKB formula is quite good, in spite of its failure to satisfy our
earlier criterion of section \ref{WKBValidity}. The accuracy of
the WKB for small fields is however due to us staying away from
the top of the barrier. We now compare equation
(\ref{AsympLimit2}) with the exact result equation
(\ref{TransmissionCefficient}) as we approach the barrier, i.e.
$V_0\rightarrow -\phi$. One mechanism for this to occur could be
due to a large image term i.e. Schottky effect, but there are
other screening effects as well as we shall see later. For
simplicity we shall not include the image effect in the
Schrodinger equation yet but merely consider the effect of
approaching the barrier top in the image free case first.  Here we
can compare the results of three formulas,the exact equation
(\ref{TransmissionCefficient}), the asymptotic formula equation
(\ref{AsympLimit1}) and the standard WKB result in which the
prefactor in equation (\ref{AsympLimit1}) is missing: these are
shown in figure ~\ref{fig 3}, where we have set $V_0 =
-0.95~\phi$. Here we see that for quite small fields $F<100 V/\mu
\rm m$, significant departures from WKB are found. These results
motivate us to further include the Schottky image term in the
Schrodinger equation which we take up in the next section.
\section{Exact solution including electron images}
\label{WithImage}
 We now turn to the quantum tunnelling problem including
the effect of the image term as in the original FN theory.  Then
equation (\ref{Schrodinger2}) becomes:
\begin{equation}
\psi^{\prime\prime}(x) + \Bigl[\epsilon+\alpha x +
\frac{\beta}{x}\Bigr]\psi(x)=0, \label{SchrodEqImage1}
\end{equation}
where the image potential is characterized by the parameter
$\beta=me^2/(2\hbar^2)$.  Transforming to the parameter $\xi$ as
before we now have the equation:
\begin{equation}
\psi^{\prime\prime}(\xi) + \Bigl[\xi
+\frac{\bar\alpha}{\xi-\bar\epsilon}\Bigr]\psi(\xi)=0,
\label{SchrodEqImage2}
\end{equation}
where $\bar\alpha=\beta{\alpha^{-1/3}}$, and $\bar \epsilon=
\epsilon{\alpha^{-2/3}}$. Clearly the image-free case
$\bar\alpha=0$ reduces to the usual Airy differential equation
equation (\ref{Schrodinger3}) and the above can be rewritten as:
\begin{equation}
  (\xi-\bar\epsilon)\psi^{\prime\prime}(\xi) + \Bigl[\xi (\xi-\bar\epsilon) + \bar\alpha\Bigr ] \psi(\xi) =
  0.
\label{SchrodEqImage3}
\end{equation}
We shall in fact further rewrite this as:
\begin{equation}
\psi^{\prime\prime}(\xi) +
\Bigl[\frac{(\xi-\lambda_1)(\xi-\lambda_2)}{(\xi-\bar\epsilon)}
\Bigr ] \psi(\xi) = 0. \label{SchrodEqImage4}
\end{equation}
in which the two roots of the quadratic are given by:
\begin{equation}
\lambda_1=\frac{1}{2}[\bar\epsilon+\sqrt{\bar\epsilon^2-4\bar\alpha}]
\quad {\rm and} \quad
\lambda_2=\frac{1}{2}[\bar\epsilon-\sqrt{\bar\epsilon^2-4\bar\alpha}].
\label{Quadratic1}
\end{equation}
Let us first discuss the WKB approximation \cite{LandauLifshitz},
which is obtained by approximating the wavefunction as:
\begin{equation}
\psi(\xi)=\frac{1}{\sqrt{p(\xi)}}\quad \exp\Bigl[i
\int^{\xi}p(\xi^\prime)d\xi^\prime + \frac{i\pi}{4}\Bigr]
\label{WKBPsi1}
\end{equation}
where:
\begin{equation}
p(\xi)=
\Bigl[\frac{(\xi-\lambda_1)(\xi-\lambda_2)}{(\xi-\bar\epsilon)}
\Bigr ]^{1/2} \label{WKBPsi2}
\end{equation}
and the lower limit in the integral in equation (\ref{WKBPsi1}) is
the smaller of $\lambda_1$ or $\lambda_2$ which depends on the
sign of $\bar\epsilon$ i.e. above or below the barrier
respectively. The evaluation of the integral in equation
(\ref{WKBPsi1}) is then a straightforward exercise in Elliptic
integral reduction which we need not repeat here
\cite{Nordheim,GoodMuller,MurphyGood}. Equation (\ref{SchrodEqImage4}) itself does not
appear to have a closed form solution in terms of known functions
although several methods of solution are available depending on
the parameters.  We postpone these to the appendices as we do not
intend to pursue detailed calculations here, but merely to
demonstrate the regimes where WKB is in need of  correction.  For
this purpose we can concentrate on the region close to the top of
the barrier at $\xi=\bar\alpha^{1/2}+\bar\epsilon$. Let us start
at the barrier top itself for which WKB predicts $T=1$ since in
this case $\upsilon(x)=0$ in equation (\ref{FNcurrent}).  Then
equation (\ref{SchrodEqImage4}) simplifies to the form:
\begin{equation}
\psi^{\prime\prime}(\xi) +
\Bigl[\frac{(\xi-\bar\epsilon/2)^2}{(\xi-\bar\epsilon)} \Bigr ]
\psi(\xi) = 0. \label{SchrodEqImageTop1}
\end{equation}
In view of the magnitude of the parameter $\bar\epsilon$ it
suffices to obtain a solution in the limit when this parameter is
large. For $\phi=1~{\rm e\rm V}$ and $100<F<1000~{\rm V}/{\rm\mu
m}$, $\bar\epsilon$ ranges from 3 to 14. In particular for the
matching conditions we are only interested in the solution near to
$\xi_0=\bar\epsilon$ which corresponds to the origin $x=0$ so that
we can transform to the variable: $y=\xi-\bar\epsilon$ which is
small for our purpose $y<<\bar\epsilon$ so that equation
(\ref{SchrodEqImageTop1}) can be approximated by:
\begin{equation}
y\psi^{\prime\prime}(y) + (y+\bar\epsilon/2)^2 \psi(y) \approx
y\psi^{\prime\prime}(y) + (\bar\epsilon/2)^2 \psi(y) = 0.
\label{SchrodEqImageTop2}
\end{equation}
The appropriate solution for the above is then given by the Hankel
function of order $1$ which upon transforming back to the original
variables $\xi$ is given by:
\begin{equation}
\psi(\xi)=e^{i\pi/4} \sqrt{\frac{\pi}{2}}\quad \bar\epsilon
(\xi-\bar\epsilon)^{1/2}
H^{(1)}_1(\bar\epsilon(\xi-\bar\epsilon)^{1/2}). \label{PsiTop1}
\end{equation}
We can now obtain the transmission coefficient as was done in the
previous section by matching the wave function and its derivative
at the origin. The result is now given by:
\begin{equation}
T=\frac{2\bar\epsilon^2 k_0^{1/3}}{|\psi(\xi_0)|^2+\epsilon^2
k_0^{1/3}+k_0^{2/3}|\psi^\prime(\xi_0)|^2} \label{TImage}
\end{equation}
where the derivative of $\psi$ is easily shown to be given in
terms of the Hankel function of order $0$:
\begin{equation}
\psi^\prime(\xi)=e^{i\pi/4} \sqrt{\frac{\pi}{8}}\quad
\bar\epsilon^2 H^{(1)}_0(\bar\epsilon(\xi-\bar\epsilon)^{1/2})
\label{PsiTopPrime}
\end{equation}
As usual we can verify that this solution satisfies unitarity
$T+R=1$ through the use of the appropriate Wronskian identities
for the Hankel functions. That this is satisfied exactly in spite
of our approximation in equation (\ref{SchrodEqImageTop2}) means
that we have picked out the essential features in the theory.
Equation (\ref{TImage}) gives zero transmission at the top of the
barrier because of a weak logarithmic singularity (typical of the
$1/x$ potential near the origin) in $\psi^\prime$. We may write
\begin{equation}
T \approx
\frac{4\pi}{\bar\epsilon^{3/2}(\ln(\delta^{1/2}\bar\epsilon))^2}
\label{TImage1}
\end{equation}
where $\delta$ is a renormalization constant, which we can
arbitrarily set to $\delta<<\bar\epsilon$. Physically, we may
think of this as correcting for the situation in which the
electron and the image charge are both at the interface by
limiting the distance of closest approach to atomic dimensions.
Equation (\ref{TImage1}) corrects for the error made by the WKB
approximation at the top of the barrier. As a convenient choice of
$\delta$, we require that it is chosen to limit the closest
approach to the origin as $x=a=1$\AA \ \footnote{By way of
comparison, the typical tunnelling length $\zeta= \frac{\phi}{e
F}$ falls in the range of $10^4$\AA\ to $10$\AA\ for $1<F<10^3$
V/$\mu$m.} so that by the definition of $\xi$:
\begin{equation}
\delta= \frac{10^{-4} F} {\phi}.
 \label{deltadefn}
\end{equation}
Before proceeding further we shall briefly mention the results
when we move slightly away from the top of the barrier.  In this
case the two roots become:
\begin{equation}
\lambda_1\approx\frac{\bar\epsilon}{2}(1+\sigma) \quad and \quad
\lambda_2\approx\frac{\bar\epsilon}{2}(1-\sigma)
\label{Quadratic2}
\end{equation}
where $\sigma=(1-4 {\bar\alpha}/{\bar\epsilon^2})^{1/2}$ and in
the limit $\bar\epsilon$ large reduces equation
(\ref{SchrodEqImageTop2}) to:
\begin{equation}
y\psi^{\prime\prime}(y) + \gamma^2 \psi(y) = 0.
\label{SchrodEqNearTop}
\end{equation}
where $\gamma=\frac{1}{2}{\bar\epsilon}(1-\sigma^2)^{1/2}$. This
poses no difficulties and the method outlined above can be adapted
to the solution. We summarize the appropriate formulas without
further ado:
\begin{equation}
T=\frac{8\bar\gamma^2 k_0^{1/3}}{|\psi(\xi_0)|^2+4\gamma^2
k_0^{1/3}+k_0^{2/3}|\psi^\prime(\xi_0)|^2} \label{TImage2}
\end{equation}
where:
\begin{equation}
\psi(\xi)=e^{i\pi/4} \sqrt{\frac{\pi}{2}}\quad 2\gamma
(\xi-\bar\epsilon)^{1/2}
H^{(1)}_1(2\gamma(\xi-\bar\epsilon)^{1/2}) \label{PsiNearTop1}
\end{equation}
and
\begin{equation}
\psi^\prime(\xi)=e^{i\pi/4} \sqrt{\frac{\pi}{8}}\quad (2\gamma)^2
H^{(1)}_0(2\gamma(\xi-\bar\epsilon)^{1/2}) \label{PsiNearTopPrime}
\end{equation}

\begin{figure}
\begin{center}
\includegraphics[scale=1.0]{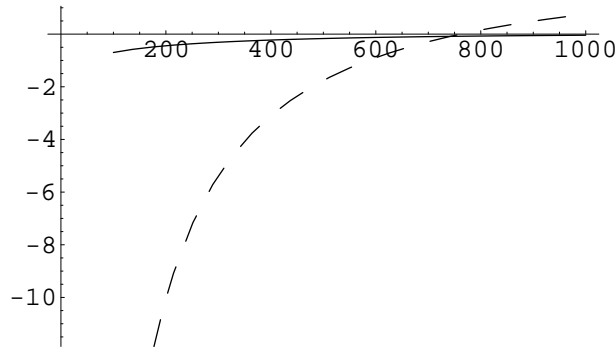} \end{center} \caption{\label{fig4}Log-linear
plot of the exact transmission coefficient equation
(\ref{TImage})(solid line) vs the FN formula (dashed line) i.e.
equation (\ref{FNcurrent}) less the prefactors in the full image
case. F is in units of $V/\mu m$.}
\end{figure}

In figure  \ref{fig4} we compare the transmission coefficient
equation (\ref{TImage}) versus the FN formula equation
(\ref{FNcurrent}) less the prefactors. We see that the FN formula
is not only in error as we trespass the barrier giving a non
physical value of $T>1$, for the region near the top of the
barrier (around $F=700 V/\mu \rm m$ here), it drops exponentially
fast as opposed to the exact solution.

\begin{figure}
\begin{center}
\includegraphics[scale=1.0]{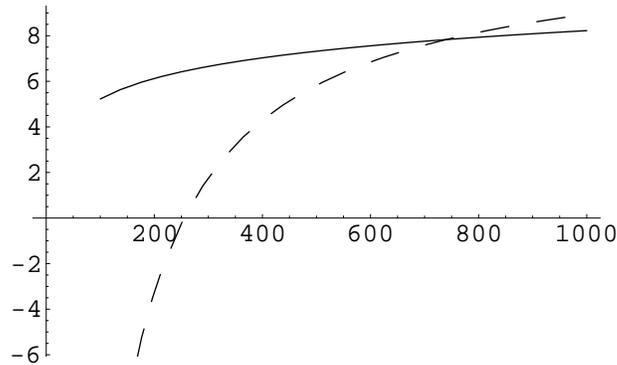} \end{center}
\caption{\label{fig5.eps}Log-linear plot of the tunnelling current
near the top of the barrier (which occurs for fields of order 700
V/$\mu \rm m$ here) (solid line) equation (\ref{CurrentNearTop2})
compared with the standard FN formula (dashed line) close to the
top of the barrier in the full image case equation
(\ref{FNcurrent}). $F$ is in units of $\rm V/\mu \rm m$.}
\end{figure}

Using equation (\ref{TImage2}) we can estimate the current near
the barrier top which occurs for $F \approx$ 700 V/$\mu$ in this
case, remembering that we have considered $\phi$=1 eV and
$W_a=2\phi$. Hence for temperatures much less than
$\epsilon_F/k_B$, the current is obtained from the integral:
\begin{equation}
J = \frac{4\pi me}{h^3}\int_{-W_a}^{\epsilon_F} T(W)(\epsilon_F-W)
dW \label{CurrentNearTop}
\end{equation}
which has to be evaluated numerically.  We can however make an
estimate using the approximate expression equation
(\ref{TImage1}). We can set the Fermi level at the top of the
barrier and integrate only over a small region of order $\delta$
near to it, since outside this region the contributions are
exponentially small. We choose $\delta$ according to equation
(\ref{deltadefn}) as before and thus the $\bar\epsilon^{-3/2}$
pre-factor dominates in the transmission coefficient since the
logarithmic term is slowly varying in the region of integration.
As we take $W_a/\phi=2$ which is large compared with $\delta$, we
now arrive at the expression for the current density as:
\begin{equation}
J \approx 0.1674\ \phi^{-3/2}\ F^{3}\quad {\rm A~cm}^{-2}
\label{CurrentNearTop2}
\end{equation}
Note that the field dependence of the current is in this case
$F^{3}$ which differs from the FN prediction of $F^2$ and could be
experimentally discernable in view of its non-exponential
dependence. We can now compare equation (\ref{CurrentNearTop2})
with the standard FN formula equation (\ref{FNcurrent}) see figure
\ref{fig5.eps}.

Having shown that we can have a large (non-exponential) current in
spite of a smaller field than in the image-free case, due to
proximity to the barrier top, we shall next examine other
mechanisms by which the barrier lowering can occur, whose
magnitude is comparable with, or might exceed the Schottky image
effect. The effect to consider is due to the dielectric screening
of the applied field F which we shall take up next.

\section{The Effects of Screening}
\label{Screening} Screening is an important property of metals or
dielectrics, and the mechanism itself plays a crucial role in the
previous section in the form of the image potential which is a
screening effect. In recent years there has been a revival of
interest in screening, \cite{Choy2000} especially in
semiconductors, where surface effects can significantly alter the
properties of devices. On the surface the screening length,
typically of the order of $\lambda_{TF}$ from Thomas-Fermi theory
ranges between the order of a few \AA \ for metals to a few $\mu
\rm m$ for semiconductors or even higher to tenths of mm due to
surface effects since it is the charge carrier density at the
surface layer that matters.  The (Laplace) screening problem for a
flat surface is easily solved for an external field $F$. This
modifies the external potential $V(x)$ in equation
(\ref{ImageFreePotential}) by having a non-zero $V_0$ such that
$V_0=-\lambda_{TF}eF$. We shall neglect an exponential penetration
field inside which in fact modifies the inner potential to
\begin{eqnarray}
V(x) &=& -W_a - \lambda_{TF}eF e^{x/\lambda_{TF}} \quad  {\rm for}
\quad x<0 - {\rm region~1} \nonumber \\
& = & -\lambda_{TF}eF-e F x \quad {\rm for} \quad  x>0 - {\rm
region~2} \label{ImageFreePotential2}
\end{eqnarray}
The latter leads to a 2D electron gas which requires a
self-consistent theory to treat since this in turn affects our
$\lambda_{TF}$ as already mentioned. Further the proper inclusion
of screening must now include its effects on the image potentials
which pick up Yukawa type terms. It suffices here to estimate the
barrier lowering from our expression for $V_0$.  Since $F$ is in
${\rm V}/{\rm\mu m}$ then a $\lambda_{TF}\approx 1~ \mu \rm m$
will completely wipe out a barrier of $1 {~\rm e\rm V}$. For 3D
this corresponds to a carrier density of order $10^{25}~{\rm
m}^{-3}$ but for a 2D layer this can be down by a $2/3$ power i.e.
$10^{17}~{\rm m}^{-2}$.  If we include the effects due to a
diminished electron affinity as in the case of diamond-like carbon
bonds, then we see that the present theory provides a strong
candidate as an explanation for the high current densities in
carbon granular samples recently observed in materials for field
emission displays \cite{PFE}, as we shall see in the next section.
\section{Relation to Experiments}
\label{Experiments} 
\begin{figure}
\begin{center}
\includegraphics[scale=1.0]{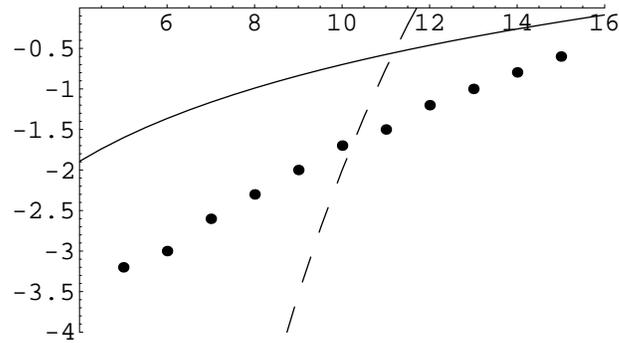} \end{center}
\caption{\label{figx.eps}Log-linear plot of the tunnelling current
vs electric field $F$, assuming that a significant barrier
lowering mechanism ($V_0=0.7\phi$), such as dielectric screening
(see text) has occurred.  The solid line is our theory, the dots
are experimental points from Tuck et al \cite{TuckLatham} while
the FN theory is given by the dash line.  In addition a large
field enhancement factor $\gamma=10$ has to be assumed for the FN
formula equation (\ref{FNcurrent}). Our result compares much more
favourably with the experimental points than does the FN theory.}
\end{figure}

Comparison with experiments such as for carbon
field emitters \cite{TuckLatham} can be made by assuming that
there is a barrier lowering mechanism such as the dielectric
screening as mentioned above.  The tunnelling current will then be
assumed to be given by equation (\ref{CurrentNearTop2}) but we
shall include a factor $\sigma$ to account for granularity or
inefficient screening (and hence barrier lowering) which
diminishes the current. This modifies equation
(\ref{CurrentNearTop2}) trivially to:
\begin{equation}
J \approx 0.1674\ \phi^{-3/2}\ \sigma^{3} F^{3}\quad \rm A\ \rm
cm^{-2} \label{FieldEmitter}
\end{equation}
In figure  \ref{figx.eps} we have plotted the results which shows
that for a granularity factor of $\sigma=0.15$ our results compare
favorably with experiments \cite{TuckLatham} bearing in mind that
our results are only estimates and not accurate numerical
evaluations\footnote{Finite temperature effects
should also be included in a more accurate calculation, providing essential
improvements to thermionic field emission theory\cite{MurphyGood,Stratton}.}.
In particular the parameters as given in the figures
have not been fine tuned in anyway.
On the contrary without assuming a significant field enhancement
due to pointed tips, the FN theory is out by numerous orders of
magnitude.  However even with the assumption of a large field
enhancement factor $\gamma=10$ which is reasonable for ellipsoids
\cite{Jaeger}, and a barrier lowering $V_0=-0.7\phi$, the FN curve
figure  \ref{figx.eps} does not in any way resemble the
experimental data, a feature we have found to occur quite commonly
in analyzing data from the carbon field emitting samples.  Note
the vast order of magnitude differences and the fast drop in the
slope of the FN curve which is shown more clearly in the inset to
figure \ref{figx.eps}.
These results clearly show that our theory has captured some
essential physics associated to tunnelling near the barrier top
which may be relevant to the carbon field emitters.

\section{Conclusion}
\label{conclusion} In conclusion we have presented a closer look
at the traditional FN theory of field emission and found that the
latter is untenable for calculations of emission currents when
significant barrier lowering occurs.  A full quantum mechanical
solution of the problem provides us with an estimate which,
coupled with a proposed barrier lowering due to screening or
negative electron affinity, constitutes a reasonable order of
magnitude estimate for the high current densities in samples with
embedded particles, not available via FN theory.  Further work is
necessary to provide more quantitative comparisons noting that the
granularity of the systems would lead to less total current
density  although on the other hand some field enhancements due to
shape effects \cite{Choy1999,Jaeger} not considered here might
partially compensate for this loss.
\section*{Acknowledgements}
 We wish to acknowledge the support of the Engineering and Physical
  Sciences Research Council through grants
 GR/M71404/01 and GR/R97047/01.

\appendix
\section{Analytical continuation of Bessel Functions}
Confusion arises with Bessel functions of order $1/3$ due to
special care needed to treat the phase factors: some popular
mathematical software uses incorrect continuation formulae.
Firstly our convention is the same as Watson's so that the
appropriate continuation formula is given by:

\begin{equation}
I_\nu(z) =\cases{e^{-i\nu\pi/2} J_\nu(z e^{i \pi/2}) &if $-\pi <
\arg(z) \le \pi/2$\cr e^{3i\nu\pi/2} J_\nu(z e^{-3 i \pi/2}) &if
$\pi/2 < \arg(z) \le \pi$. \cr} \label{Watson1}
\end{equation}

An alternative approach is to perform the continuation for
negative $z$ as:
\def\imag{i}
\begin{eqnarray}
 u_1&=&\frac{{\sqrt{\pi }}\,{\sqrt{x}}\,
     J_{-1/3}(
      \frac{2}{3}\,x^{\frac{3}{2}})}{3}
   = \frac{\imag }{3}\,{\sqrt{\pi }}\,
   {\sqrt{\vert x\vert }}\,J_{-1/3}(
    \frac{-2\,\imag }{3}\,
     \vert x\vert ^{\frac{3}{2}}) \nonumber \\
     &=&
  \frac{\imag }{3}\,
   e^{\imag\,\pi/6 }\,
   {\sqrt{\pi }}\,{\sqrt{\vert x\vert }}\,
   I_{-1/3}(
    \frac{2\,\vert x\vert^{\frac{3}{2}}}{3})\label{eq_error},
\end{eqnarray}
which is incorrect.  The negative argument prevents the use of the
first form of equation(\ref{Watson1}).  Instead the branch cut
properties requires the continuation formula \cite{Watson}:
\begin{equation}
J_\nu(e^{i m\pi/2} z) = e^{i m\pi/2} J_\nu( z)
\end{equation}
which holds similarly for the $I_{\nu}$ functions. Then:
\begin{equation}
\fl
 u_1=\frac{{\sqrt{\pi }}\,{\sqrt{x}}\,
     J_{-1/3}(
      \frac{2}{3}\,x^{\frac{3}{2}})}{3}
   = \frac{\imag }{3}\,{\sqrt{\pi }}\,
   {\sqrt{\vert x\vert }}\,J_{-1/3}(
    \frac{-2\,\imag }{3}\,
     \vert x\vert ^{\frac{3}{2}}) =
   \frac{1}{3} {\sqrt{\pi }}\,{\sqrt{\vert x\vert }}\,
   I_{-1/3}(
    \frac{2\,\vert x\vert^{\frac{3}{2}}}{3}),
\end{equation}
without any complex phase factors.  Similarly $u_2$ picks up a
minus sign:
\begin{equation}
\fl
 u_2=\frac{{\sqrt{\pi }}\,{\sqrt{x}}\,
     J_{1/3}(
      \frac{2}{3}\,x^{\frac{3}{2}})}{3}
   = \frac{\imag }{3}\,{\sqrt{\pi }}\,
   {\sqrt{\vert x\vert }}\,J_{1/3}(
    \frac{-2\,\imag }{3}\,
     \vert x\vert ^{\frac{3}{2}}) = \frac{1}{3}
   - {\sqrt{\pi }}\,{\sqrt{\vert x\vert }}\,
   I_{1/3}(
    \frac{2\,\vert x\vert^{\frac{3}{2}}}{3}).
\end{equation}
Hence one has to manually continue the $J_{\nu}$ to the $I_{\nu}$
functions for negative arguments when if using the incorrect
continuation, equation (\ref{eq_error}). We have found that
Mathematica v4.1 uses equation (\ref{eq_error}).

\section{Power series solution for the full image problem I}
The generic form of the full image problem is given by the
following second order differential equation
\begin{equation}
  \psi^{\prime\prime} + \Bigl( \epsilon + \alpha x + \frac{\beta}{x} \Bigr) \psi = 0.
\label{eqnA1}
\end{equation}
As before a straightforward linear transformation: $\xi =
(x+\frac{\epsilon}{\alpha})\alpha^\frac{1}{3}$ leads to the
following form:
\begin{equation}
  \psi^{\prime\prime} + \xi \psi + \frac{\bar\alpha}{\xi-\bar\epsilon} \psi =
  0,
\label{eqnA2}
\end{equation}
where $\bar\alpha=\beta\alpha^{-\frac{1}{3}}$, and $\bar \epsilon=
\epsilon{\alpha^{-2/3}}$. Clearly the image free case
$\bar\alpha=0$ reduces to the usual Airy differential equation.
Equation (\ref{eqnA2}) can be rewritten as:
\begin{equation}
  (\xi-\bar\epsilon)\psi^{\prime\prime} + \xi (\xi-\bar\epsilon) \psi + \bar\alpha \psi =
  0.
\label{eqnA3}
\end{equation}
Unfortunately this equation does not belong to the standard
hypergeometric class and cannot therefore be treated by the usual
mathematical physics functions.   However using the standard
Frobenius method one can obtain the following power series solutions.
Let
\begin{equation}
  u_m(\xi) = \sum_{n=0}^{\infty}a_n \xi^{m+n},
\label{eqnA4}
\end{equation}
then equation (\ref{eqnA4}) leads to the following 5 term
recursion relation for the determination of the coefficients
$a_n$:
\begin{eqnarray}
a_{n-1} (m+n-1)(m+n-2) &-& a_n \bar\epsilon (m+n)(m+n-1) \nonumber
\\ &+& a_{n-4}
 - \bar\epsilon a_{n-3} + \bar \alpha a_{n-2} = 0.
\label{eqnA5}
\end{eqnarray}
All the coefficients are well determined in terms of the
coefficient $a_0$. The first term satisfies
\begin{equation}
 \bar\epsilon m (m-1) a_0 = 0,
\label{eqnA6}
\end{equation}
giving $m = 0\ or\ m=1$ which shows as expected that we shall
have two linearly independent solutions. Carrying on we have
$a_1=0$ and
\begin{equation}
 a_2 = \frac{\bar\alpha}{\bar\epsilon} \frac{a_0}{(m+2)(m+1)}.
\label{eqnA7}
\end{equation}
Notice that $a_1=0$ regardless but $a_2$ is only finite for
$\bar\alpha\ne 0,$ i.e. $\beta \ne 0$.
The following coefficients can be sequentially determined:
\begin{eqnarray}
 a_3 &=& \frac{a_2(m+1)}{\bar\epsilon (m+3)} -
 \frac{a_0}{(m+3)(m+2)},
\label{eqnA8}\\
 a_4 &=& \frac{a_3(m+2)}{\bar\epsilon (m+4)} +
 \frac{a_0}{\bar\epsilon (m+4)(m+3)} + \frac{\bar\alpha}{\bar\epsilon} \frac{a_2}{(m+4)(m+3)},
\label{eqnA9}
\end{eqnarray}
and so on.  The general 5 term recursion relation can be
re-written as
\begin{eqnarray}
 a_n &=& \frac{1}{(m+n)(m+n-1)}\nonumber \\
 & &\Bigl[
 \frac{1}{\bar\epsilon} a_{n-1}(m+n-1)(m+n-2) + \frac{\bar\alpha}{\bar\epsilon} a_{n-2} - a_{n-3} + \frac{1}{\bar\epsilon}
 a_{n-4}\Bigr ].
\label{eqnA10}
\end{eqnarray}
Unfortunately there is no way to simplify this any further. When
$\bar\alpha=0$ (image free case) the above in fact reduces to a
two term relation given by:
\begin{equation}
a_n= - \frac{a_{n-3}}{(m+n)(m+n-1)}. \label{eqnA11}
\end{equation}
This immediately leads us to the two independent solutions, for
$m=0$
\begin{eqnarray}
u_0(\xi) &=& 1 + \sum_{n=1}^\infty \frac{(-1)^n
1.4.7\dots(3n-2)}{(3n)!} \xi^{3n} \nonumber \\
&=& \frac{\Gamma(2/3)}{3^{1/3}}\xi^{1/2}
J_{-1/3}(\frac{2}{3}\xi^{3/2}), \label{eqnA12}
\end{eqnarray}
and for $m=1$
\begin{eqnarray}
u_1(\xi) &=& \xi + \sum_{n=1}^\infty \frac{(-1)^n
2.5.8\dots(3n-1)}{(3n+1)!} \xi^{3n+1} \nonumber \\
&=& {3^{1/3}\Gamma(4/3)} \xi^{1/2} J_{1/3}(\frac{2}{3}\xi^{3/2}),
\label{eqnA13}
\end{eqnarray}
which are the series expansions for the Bessel functions of order
$1/3$.  The oscillatory nature of these functions means that the
convergence is in general poor and some more sophisticated
re-summation method must be used to obtain in particular the
asymptotic properties of these functions.  In the next appendix we
shall show just one method as to how this can be achieved.
\section{Power series solution for the full image problem II}
An alternative solution for the full image problem which is an
effective re-summation of the previous power series solution that
will allow us to perform numerical calculations, will now be
shown. The approach is to first use a Laplace transform on
equation (\ref{eqnA1}). Let
\begin{equation}
 \psi(\xi)=\int_{C} e^{\xi t} f(t) dt,
\label{eqnB1}
\end{equation}
over some contour $C$, then the resulting differential equation in
$\psi$ becomes converted into a second-order differential equation
in $f$:
\begin{equation}
 f^{\prime\prime} -(t^2-\bar\epsilon)f^\prime+(\bar\alpha-\bar\epsilon\ t^2-2t)f=0.
\label{eqnB2}
\end{equation}
It is easy to show that the solution for $\bar\alpha = 0$ reduces
to the Airy integral where $f(t)=e^{t^3/3}$ so we shall factorize
$f$ in the form: $f=ug$ where $u=e^{t^3/3}$.  The equation for $g$
now simplifies tremendously:
\begin{equation}
 g^{\prime\prime} +(t^2+\bar\epsilon)g^\prime+\bar\alpha g=0.
\label{eqnB3}
\end{equation}
Unfortunately there is still no closed form solution for this
equation although it can be proved that solutions for $g$ must be
an entire function, by which we can develop an expansion as an
infinite series:
\begin{equation}
 g = \sum_{n=0}^{\infty} a_n t^{m+n}.
\label{eqnB4}
\end{equation}
The boundary condition requires that when $\bar\alpha \rightarrow
0$ then $g\rightarrow 1$.  With this we can show that $m=0$ is the
only allowed index.  Then we have the following 4 term recursion
relation:
\begin{equation}
(n-1)(n a_n+\bar\epsilon a_{n-1}) +(n-3)a_{n-3}+\bar\alpha a_{n-2}
= 0. \label{eqnB5}
\end{equation}
The first few solutions are easily obtained: $a_0=1$,
$a_2=\frac{\bar\alpha}{2}$,\
$a_3=\frac{\bar\epsilon\bar\alpha}{6}$,\
$a_4=\frac{\bar\alpha}{24}(\bar\alpha-\bar\epsilon^2)$ and so on.
As required only the first term survives in the limit $\bar\alpha
= 0$.  Interestingly the 4 term relation equation (\ref{eqnB5})
can actually be reduced to a three term one. Let
$\gamma_n=\frac{a_n}{a_{n-1}}$ then
\begin{equation}
(n-1)(n \gamma_n+\bar\epsilon)\gamma_{n-1}
+\frac{(n-3)}{\gamma_{n-2}}+\bar\alpha = 0, \label{eqnB6}
\end{equation}
which may be more convenient for computation. Finally we have now
effectively re-summed the series to obtain
\begin{equation}
 \psi(\xi)=\sum_{n=0}^\infty a_n \int_C t^n e^{\xi t + \frac{1}{3} t^3} dt.
 \label{eqnB7}
\end{equation}
We now see that
\begin{eqnarray}
 \int_C t^n e^{\xi t + \frac{1}{3} t^3} dt &=&\frac{d^n}{d\xi^n} \int_C e^{\xi t + \frac{1}{3}
 t^3}
dt  \nonumber \\
 &=& \frac{d^n}{d\xi^n}\Bigl[\xi^{1/2}Z_{\pm 1/3}(\frac{2}{3}\xi^{3/2})\Bigr],
\label{eqnB8}
\end{eqnarray}
the last term following from the equivalence between the Airy
functions and the Bessel functions of order $1/3$.  In view of the
recursion properties for the derivatives of the Bessel functions,
we have now converted the solution effectively into an infinite
series of Bessel functions. Only the first term survives in the
image free case which is now trivial and all the oscillatory bits
are now absorbed into the Bessel functions.
\end{document}